\documentclass{PoS}

\font\tenrsfs=rsfs10 at 12pt
\font\sevenrsfs=rsfs7
\font\fiversfs=rsfs5
\newfam\rsfsfam
\textfont\rsfsfam=\tenrsfs
\scriptfont\rsfsfam=\sevenrsfs
\scriptscriptfont\rsfsfam=\fiversfs
\def\mathscr#1{{\fam\rsfsfam\relax#1}}

\makeatletter
\def\art{\@ifnextchar[{\eart}{\oart}}
\def\eart[#1]#2#3#4#5#6{{\rm #2}, {#3 #4} {\rm (#6) #5} [{{#1}}]}
\def\hepart[#1]#2{{\rm #2, {#1}}}
\newcommand{\oart}[5]{{\rm #1}, {#2 #3} {\rm (#5) #4}}

\def\circa#1{\,\raise.3ex\hbox{$#1$\kern-.75em\lower1ex\hbox{$\sim$}}\,}

\title{Minimal Dark Matter predictions and the PAMELA positron excess}

\ShortTitle{Minimal Dark Matter and PAMELA}

\author{\speaker{Marco Cirelli} \thanks{We are indebted to Gianfranco Bertone, Torsten Bringmann, Pierre Brun and Nicolao Fornengo for very useful discussions and we thank the organizers of the idm08 conference for the very stimulating atmosphere. The work of M.C. is partially supported by the french ANR grants DARKPHYS and PHYS@COL\&COS. We thank the EU Marie Curie Research \& Training network "UniverseNet" (MRTN-CT-2006-035863) for support.}\\
        Institut de Physique Th\'eorique, CNRS, URA 2306 \& CEA/Saclay,\\ 
	F-91191 Gif-sur-Yvette, France\\
        E-mail: \email{marco.cirelli@cea.fr}}

\author{Alessandro Strumia\\
        Dipartimento di Fisica dell'Universit{\`a} di Pisa \& INFN, Italia\\
        E-mail: \email{astrumia@df.unipi.it}}

\abstract{We present Minimal Dark Matter and its univocal predictions for Dark Matter observables.
During the idm08 conference, PAMELA presented preliminary results showing
an anomaly in the positron fraction: we find a good agreement, with a modest astrophysical boost factor.}

\FullConference{Identification of dark matter 2008\\
		 August 18-22, 2008\\
		 Stockholm, Sweden\\
		 --\\
		 Preprint number: SACLAY--t08/163}

\begin{document}

\section{Introduction}
%%%%%%%%%%%%%%%%%%%%%%%%%%%%%%%%%%%%%%%%%%%%%%%%%%%%%%%%%%%%%%%%%%%%%%

The quest for the identification of the missing mass of the universe has been with us since many decades now~\cite{sum}. While explanations in terms of modifications of Newtonian gravity or General Relativity become more and more contrived, 
evidence for the particle nature of such Dark Matter (DM) now comes from many astrophysical and cosmological observations. 
Non-baryonic new particles that may fulfill the r\^ole of DM have emerged in the latest decades within many 
Beyond the Standard Model (SM) theories, most notably supersymmetry. 
These constructions try to naturally explain the hierarchy between the ElectroWeak (EW) scale and the Planck scale and, in doing so, introduce a host of new particles with EW masses and interactions. 
Some of these particles can be good DM candidates (e.g.\ the lightest neutralino). 
DM stability is the result of extra features introduced by hand (e.g.\ R-parity), 
usually necessary also to recover many good properties of the SM that are lost in these extensions
(automatic conservation of baryon number, lepton number, etc).
Finally, the richness of these theories implies the introduction of many unknown new parameters (e.g.\ all sparticle masses), so that the phenomenology of the DM candidate remains unclear.

The Minimal Dark Matter (MDM) proposal~\cite{MDM} originates from different motivations: 
focussing on the Dark Matter problem only, we add to the SM the minimal amount of new physics (just one extra EW multiplet $\chi$) and search for the minimal assignments of its quantum numbers
(spin, isospin and hypercharge) that make it a Dark Matter candidate without ad hoc extra features, and without ruining the good features of the SM.
As detailed in the following section, we do find one optimal candidate, and we here focus on it.
Its only free parameter (the DM mass) is fixed from  the cosmological DM abundance, so that
any DM observable can be univocally predicted.

Indirect searches are one of the most promising ways to detect Dark Matter. DM particles in the galactic halo are expected to annihilate and produce fluxes of cosmic rays that propagate through the galaxy and reach the earth.
Their energy spectra carry important information on the nature of the DM particle (mass and primary annihilation channels). Many experiments searched  for signatures of DM annihilations in the fluxes of $\gamma$ rays, positrons and antiprotons.

At the idm08 conference, the PAMELA experiment~\cite{PAMELA} reported preliminary results 
that seem to be the first strong hint for a DM indirect signal.
We here assume that PAMELA data will be confirmed and that
on-going re-evaluations of the astrophysical backgrounds will confirm previous studies, such that
the PAMELA excess implies WIMP DM (non-WIMP DM candidates such as the gravitino may remain viable if unstable~\cite{Ibarra}; DM candidates with a relic density due to a baryon-like asymmetry are disfavored).
In view of its univocal predictions, MDM could have been immediately excluded, rendering this talk unnecessary.
After introducing the MDM model we compare its predictions, as previously computed in~\cite{MDMindirect},
with the PAMELA results.

\section{Minimal Dark Matter}
The MDM model is constructed  by adding on top of the SM a single multiplet $\chi \oplus\bar\chi$ with weak interactions, fully determined by its hypercharge $Y$ and by the number of its SU$(2)_L$ components,
$n=\{2,3,4,5,\ldots\}$. The Lagrangian is therefore `minimal':
$$
\mathscr{L} = \mathscr{L}_{\rm SM} + \frac{1}{2}
\left\{\begin{array}{ll}
 \bar{\chi} (i D\hspace{-1.4ex}/\hspace{0.5ex}+M) \chi & \hbox{for fermionic $\chi$}\\
|D_\mu \chi|^2 - M^2 |\chi|^2& \hbox{for scalar $\chi$}
\end{array}\right.
$$
where the gauge-covariant derivative $D_\mu$ contains the usual electroweak gauge couplings and vectors, $M$ is a tree level mass term (the only free parameter).
%and $c = 1/2$ for a real scalar or a Majorana fermion and $c = 1$ for a complex scalar or a Dirac fermion gives the canonical normalization. 
Any additional term (such as Yukawa couplings with SM fields) will be forbidden by gauge and Lorentz invariance, as detailed below. 
%We try to choose $n$ and $Y$ such that any other additional term is forbidden by gauge and Lorentz invariance,
%so that DM is automatically stable.

For a given assignment of $n$ there are few assignments of the hypercharge $Y$ such that one
component of the $\chi$ multiplet has electric charge $Q=T_3+Y=0$, as needed for a DM candidate.
For instance, for $n=2$, since $T_3 = \pm 1/2$, the only possibility is $Y=\mp 1/2$.
For $n=5$ one can have $Y = \{ 0, \pm 1, \pm 2\}$, and so on.

\smallskip

But MDM candidates with $Y \neq 0$ interact with the nuclei of direct detection experiments via exchange of a $Z$ boson, giving rise to an effect not seen by the Xenon and CDMS~\cite{dir} experiments.
%On a nucleus ${\cal N}$ of mass $M_{\cal N}$ and atomic (mass) number $Z$ ($N$), the spin-independent elastic cross section $\sigma(DM {\cal N} \to DM {\cal N}) = c/2\pi \ G_{\rm F}^2 M_{\cal N}^2 Y^2 (N -(1- 4 \sin^2\theta_w)Z )^2$~\cite{GoodmanWitten}, much larger than the current bounds from Xenon~\cite{Xenon} and CDMS\cite{CDMS} (here $c = 1 (4)$ for fermions (scalars)). 
Thus we restrict to candidates with $Y=0$, and therefore to odd $n$ multiplets.
Also, the list of possible MDM candidates has to stop at $n \le 5~ (8)$ for fermions (scalars) because larger multiplets would cause the running of $g_2$ to hit a Landau pole below the Planck scale.

Next we inspect which remaining candidates are stable against decay into SM particles. For instance, the fermionic 3-plet with hypercharge $Y=0$ would couple with a Yukawa operator $\chi L H$ with a SM lepton doublet $L$ and a Higgs field $H$ and decay. This is not a viable DM candidate, unless the operator is eliminated by some ad hoc symmetry.
For another instance, the scalar 5-plet with $Y=0$ would couple to four Higgs fields with a dimension 5 operator $\chi HHH^*H^*/M_{\rm Pl}$, suppressed by one power of the Planck scale. Despite the suppression, the resulting life-time is shorter than the age of the universe, so that this is not a viable DM candidate.
Now, the crucial observation is that, given the known SM particle content, there are multiplets  that cannot couple to SM fields and are therefore automatically stable DM candidates. Only two possibilities emerge:
a $n=5$ fermion, or a $n=7$ scalar.
But since the latter may have non-minimal quartic couplings with the Higgs field, we will set it aside and focus here on the former for minimality. 

Quantum corrections due to a loop of gauge bosons generate a small mass splitting between the components of $\chi$. The lightest component turns out to be the neutral one (as required by DM phenomenology), and the $Q=\pm1$ partners are $166\,{\rm MeV}$ heavier~\cite{MDM}.

We can now compute the DM cosmological abundance  as a function of the only free parameter, the mass $M$. 
The abundance measured by cosmology, $\Omega_{\rm DM} h^2 =0.110 \pm 0.005$~\cite{cosmoDM},
is matched for $M=(9.6\pm0.2)$ TeV~\cite{MDMastro}.  
This result is obtained solving the relevant Boltzmann equation
taking into account all co-annihilations and, importantly, electroweak Sommerfeld corrections~\cite{Hisano}.
This non-perturbative phenomenon significantly enhances
non-relativistic annihilations of particles with mass $M\circa{>}M_W/\alpha_2$.
As a result the (co)-annihilation cross section
$\sigma v$ grows as $v\to0$, so that astrophysical signals ($v\sim 10^{-3}$ in our galaxy),
being much more enhanced than DM annihilations in cosmology ($v\sim 0.2$ at freeze-out),
are detectably large despite the large multi-TeV DM mass $M$.

Elastic scattering of $\chi$ on nuclei occurs at 1-loop via the exchange of $W$'s and Higgs~\cite{MDM, MDMastro},
giving rise to a negligible spin-dependent cross section and to a
spin-independent cross section $\sigma_{\rm SI}({\rm DM} \ {\rm N}) \approx 10^{-44} {\rm cm}^2$ 
(up to reducible uncertainties due to QCD and to the unknown Higgs mass), within the reach of the next generation of direct detection experiments~\cite{dir}.

In summary, the MDM construction singles out a 
$$\hbox{fermionic SU(2)$_L$ 5-plet with zero hypercharge}$$
as providing a fully viable, automatically stable DM candidate. It is called `Minimal Dark Matter' since it is described by the minimal gauge-covariant Lagrangian. Its mass is fixed at $(9.6\pm 0.2)$ TeV and its phenomenology is fully computable with no free particle-physics parameters.

\section{Indirect signatures and the PAMELA positron excess}
\begin{figure}[t]
\vspace{-0.64cm}
\begin{center}
\includegraphics[width=\textwidth]{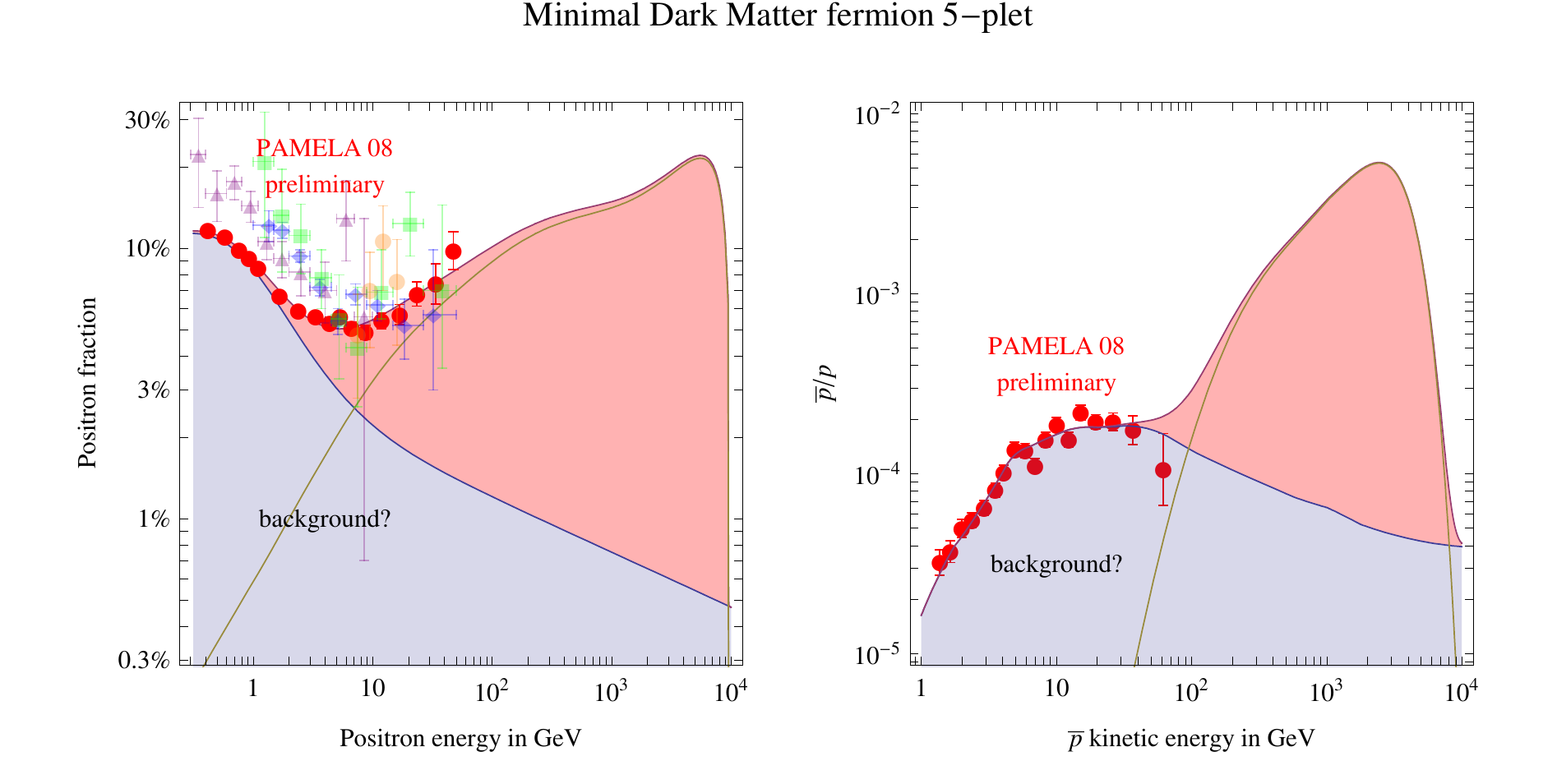}
\vspace{-0.62cm}
\caption{\em\label{fig:PAMELA} The PAMELA preliminary data~\cite{PAMELA} compared with the fermion 5-plet MDM prediction,
at the best-fit point for the astrophysical parameters.}
\end{center}
\vspace{-0.89cm}
\end{figure}

The MDM fermionic 5-plet annihilates at tree level into $W^+W^-$, and into $\gamma \gamma$, $\gamma Z$ and $ZZ$ at one loop. We neglected 3-body primary final states.
The annihilation cross sections at $v\sim 10^{-3}$ are large thanks to the Sommerfeld enhancement:  
for $M=9.6$ TeV one has
$$
\langle \sigma v \rangle_{WW} =  1.1\cdot 10^{-23}\ \frac{{\rm cm}^3}{{\rm sec}} ,\qquad
\langle \sigma v \rangle_{\gamma\gamma} =  \langle \sigma v \rangle_{\gamma Z} \frac{\tan^2\theta_w}{2} = \langle \sigma v \rangle_{ZZ} \tan^4\theta_w = 3 \cdot 10^{-25}\ \frac{{\rm cm}^3}{{\rm sec}}.
$$
The Sommerfeld corrections also introduce a strong dependence on $M$, such that, within its $3\sigma$ range, the cross sections change by one order of magnitude around these central values.
The resulting spectra of $e^+$ and $\bar p$, plotted in fig.~\ref{fig:PAMELA}, are obtained from the primary spectra computed 
taking into account spin-correlations and propagated in the galactic halo~\cite{MDMindirect}.

The PAMELA experiment presented preliminary results~\cite{PAMELA} for the fluxes of antiprotons and positrons in the cosmic rays. The latters show an excess at $E_{e^+} = (10 - 60)$ GeV with respect to the expected background, compatibly with hints that previous experiments (e.g. HEAT) had already suggested with a much lower significance. At the same time, the $\bar p$ data show no anomaly.

\smallskip

We tried to perform a preliminary fit of the preliminary PAMELA data to have a feeling of which set of astrophysical assumptions allows to reproduce the data and how well. 
We have taken the $e^+$ and $\bar p$ astrophysical backgrounds from~\cite{bck}, and multiplied each one of them
times a free normalization factor and times a spectral correction $E^p$ with $p=0\pm0.1$. This conservatively mimics the estimated uncertainties.
Concerning the DM signal, we smoothly vary between the possible halo models and between the propagation configurations considered in~\cite{MDMindirect}, assuming that they are the same for $e^+$ and $\bar p$:
this should reasonably approximate a precise fit where galactic parameters are extracted from CR data.
Uncertainties on $\bar p$ propagation mainly affect the overall $\bar p$ flux, and we anyway
assume different energy-independent boost factors $B_e$ and $B_p$ for $\bar p$ and $e^+$.
Since the overall normalization of the signal is anyhow uncertain,
we presume that it is safe to neglect possible statistical correlations among the PAMELA data points.

Under these assumptions, the best MDM fit is at $B_e \cdot \sigma v = 4~10^{-22}\,{\rm cm}^3/{\rm sec}$ 
(i.e.\ $3\circa{<}B_e \circa{<} 100$) and for a propagation model intermediate between `MED' and `MAX';
the halo model is not significantly constrained. 
Fig.~\ref{fig:PAMELA} shows the MDM fit superimposed to the preliminary PAMELA data; we here used $B_p = 3$, and this fit does not significantly deteriorate until much larger values.
For the moment uncertainties can only be estimated, so that the fact that this fit has $\chi^2/{\rm dof} \sim 1$ is encouraging but cannot be taken as an overall quality indicator.
Alternative tools can be employed.

%First,  fitting the positron excess with a toy flux proportional to $\Phi_{e^+}\propto E^p$, our fit
%suggests that PAMELA data significantly constrain $p$ to be $p=-2.64\pm 0.06$.
%% I checked that this 0.06 is real and does not come because of the assumed uncert. on the slope of the bck
%This value happens to be compatible with the slope predicted by MDM,
%$-2.1< d\ln\Phi_{e^+}/d\ln E< -2.9$ 
%(a range obtained varying the astrophysical parameters and
%the energy $E=(20-60)\,{\rm GeV}$).\footnote{We emphasize that this
%slope is mostly determined by the shape of the spectrum from $W$ fragmentation.
%This can be contrasted to the alternative scenario of DM DM $\to e^+e^-$ annihilations:
%positron propagation in the galaxy transforms the
%positron line at $E = M$ into a continuum with $E<M$:
%in this case the slope is determined by uncertain astrophysics.}

We varied $M$ in order to see if the MDM value $M\sim 10\,{\rm TeV}$ is preferred by data. We find that increasing the DM mass  above $10\,{\rm TeV}$  starts to give a poorer fit of the $e^+$ spectrum.
Lowering the DM mass, one needs to increasingly reduce free parameters such as $B_p/B_e$ in order
to generate the $e^+$ excess without giving at the same time an unseen $\bar p$ excess.

The $e^+$ and $\bar p$ spectra  will be measured by
PAMELA  (possibly up to 270 GeV for $e^+$ and 150 GeV for $\bar p$)  and later by AMS-08 (up to about 1 TeV).
MDM predicts that the positron fraction should continue to grow, and that
an anomaly should appear in the $\bar p$ spectrum, unless
$\bar{p}$ have an unfavorable boost factor or propagation in our galaxy.

Collateral constraints must be considered. The $e^\pm$ from DM annihilations lead to a synchrotron radiation~\cite{MDMindirect} at the level of `WMAP haze' anomaly~\cite{WMAPhaze}. 
Ref.~\cite{Ullio} claims that very strong bounds on the DM annihilation cross section can be inferred 
from infrared and X-ray observations of the galactic center region, modeled assuming
a certain magnetic field and DM density, that gets extremely high close to the central black hole
leading to a high rate of DM annihilations.
In this region DM becomes relativistic, and in the MDM case this means that the
Sommerfeld enhancement disappears, leaving a small  annihilation cross section,
$\sigma \sim  \alpha_2^2/M^2\sim 10^{-28}\,{\rm cm}^3/{\rm sec}$ that would not contradict the strong bounds of~\cite{Ullio}. A dedicated computation of the MDM prediction together with a precise description of the galactic center 
is necessary to quantitatively clarify this issue.

\smallskip

To conclude: we presented Minimal Dark Matter. Like string theory, MDM has no free parameters, and thereby makes univocal predictions, falsifiable by any single experimental result.
The preliminary data from PAMELA, presented during idm08, show an excess in the flux of cosmic ray positrons at 10-60 GeV which matches the MDM prediction.
Let us compare with supersymmetry, the theoretically favored scenario: 
slepton masses can be fine-tuned to be quasi-degenerate with the lightest neutralino in order to enhance 3-body annihilations obtaining the correct relic abundance and a $e^+$ spectrum that, with a boost factor of $\circa{>}10^4$, can be compatible with the PAMELA excess~\cite{BBE}: in such a case the $e^+$ fraction should decrease at higher energy. MDM predicts the continuing rise of fig.~1a.
The PAMELA results recently published on the arXiv~\cite{PAMELA} have one extra data-point at 80 GeV, still consistent with MDM predictions~\cite{MDMindirect}.
The nearby pulsars Geminga or B0656+14 could also produce a rising $e^+$ fraction, together with an angular anisotropy~\cite{Geminga}.

%Upcoming data at higher energies from PAMELA will hopefully clarify the issue.

\end{document}